# Is the friction angle the maximum slope of a free surface of a non cohesive material?

**A. Modaressi & P. Evesque**
Lab MSSMat, UMR 8579 CNRS, Ecole Centrale Paris
92295 CHATENAY-MALABRY, France, e-mail evesque@mssmat.ecp.fr

**Abstract:**
*Starting from a symmetric triangular pile with a horizontal basis and rotating the basis in the vertical plane, we have determined the evolution of the stress distribution as a function of the basis inclination using Finite Elements method with an elastic-perfectly plastic constitutive model, defined by its friction angle $\varphi$ , without cohesion. It is found that when the yield function is the Drücker-Prager one, stress distribution satisfying equilibrium can be found even when one of the free-surface slopes $\theta$ is larger than $\varphi$. This means that piles with $\theta>\varphi$ can be (at least) marginally stable and that slope rotation is not always a destabilising perturbation direction. On the contrary, it is found that $\theta$ cannot overpass $\varphi$ when a Mohr-Coulomb yield function is used. Theoretical explanation of these facts is given which enlightens the role plaid by the intermediate principal stress $\sigma_2$ in both cases of the Mohr-Coulomb criterion and of the Drücker-Prager one.*
*It is then argued that the Mohr-Coulomb criterion assumes a spontaneous symmetry breaking,as soon as $\sigma_2\neq\sigma_3$;this is not physical most likely; so this criterion shall be replaced by some Drücker-Prager criterion in the vicinity of $\sigma_2=\sigma_3$; as this Drücker-Prager criterion leads to some anomalous friction behaviour, since the present work demonstrates that the slope $\theta$ of a pile obeying this modelling can be larger than the friction angle $\varphi$ , these numerical computations enlighten the avalanche process: they show that no dynamical angle $\varphi_{dyn}<\varphi$ is needed to understand avalanching. It is in agreement with previous experimental results. Furthermore, these results show that the maximum angle of repose $\theta_M$ can be modified using cyclic rotations; we propose then a procedure which allows to achieve $\theta_M=\varphi$ .*

---

A great number of works has been dealing with the stress distribution under conic or triangular piles recently and the problem of the stability of their free surface has been investigated intensively.

In order to describe the avalanching process, the models [1] introduce a maximum angle of repose $\theta_M$ and a dynamical friction angle $\varphi_{dyn}$, such as $\theta_M>\varphi_{dyn}$. It is considered that $\theta_M$ looks like the static friction and that this static angle may depend on the history of the stress, on the time during which the system was in equilibrium,….

On the other hand classical experiments of soil mechanics have identified a quasi-static regime for which time is not a parameter; it is characterised by a well defined rheology whose parameters do not depend on the rate of deformation; it depends on a unique friction angle $\varphi$ that governs the mechanics at large deformation; within this approach $\varphi$ is independent of the stress field and of the initial specific volume, although the rheology of the medium itself depends on the specific volume at small deformation. Using avalanche experiment in centrifuge, we have also demonstrated that the maximum angle of repose depends on the pile density and on gravity as soil mechanics predicts; furthermore the dynamic friction angle measured [2-5] is equal to the value of the quasi-static one, measured with a triaxial test, *cf.* Fig.1.

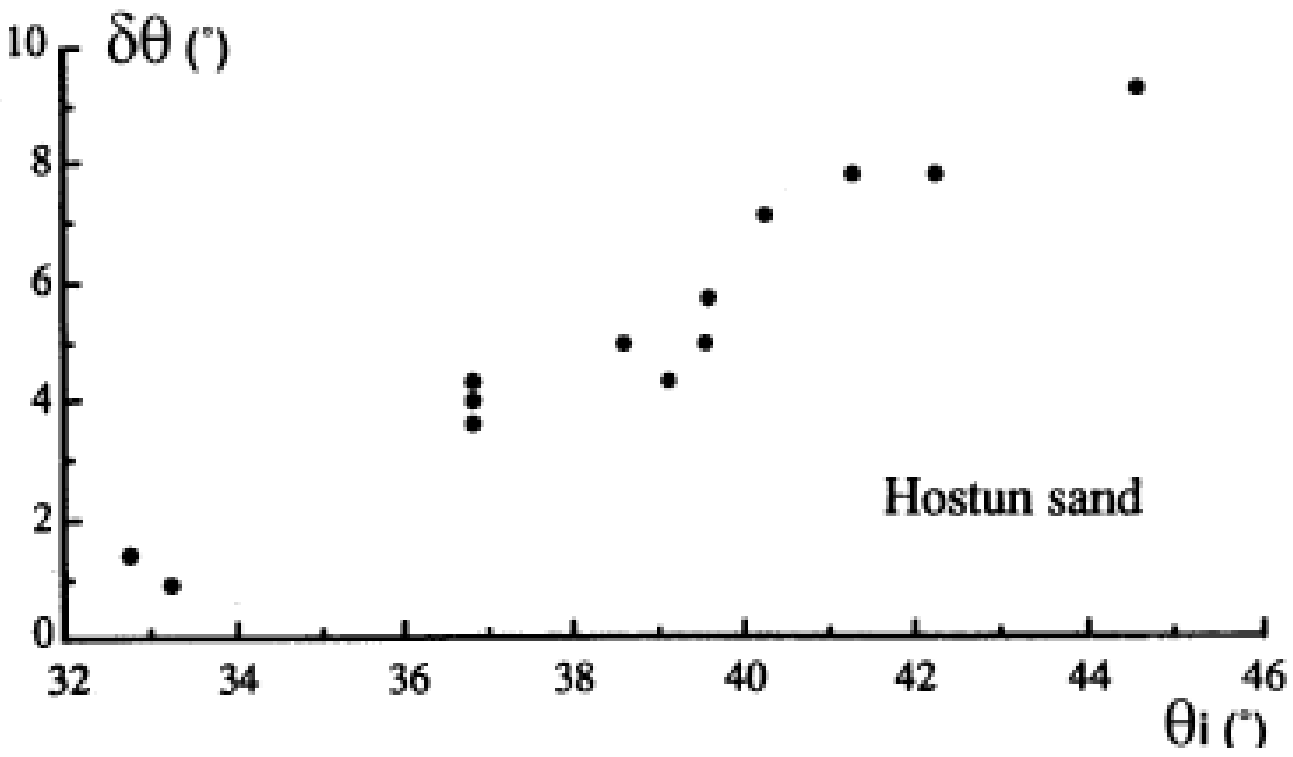

***Figure 1.*** *Avalanche size* $\delta\theta$ *vs. maximum angle of repose* $\theta_i$, *for avalanches generated in centrifuge by rotating a box and letting the sand flow outside the box. The slope of* $\delta\theta$ *vs.* $\theta_i$ *is equal to* 1, *and* $\theta_i=\varphi_{dyn}$ *is achieved when* $\delta\theta=0$. *Hence, flows stop always at* $\theta_e=\varphi$, *. It is found also that* $\varphi$ *corresponds also to the friction angle measured with triaxial test.*

So, a question arises then: is it really needed to define a dynamical friction angle different from the quasi-static one? What is the physical process which governs the maximum angle of repose $\theta_M$. Is this physical process already contained in the soil mechanics approach? Or shall we introduce a new effect?

In the same way, there has been a recent debate about the stress distribution in a conic pile, for which one can find experimentally a dip of vertical stress just in the middle of the pile, depending on the building process. It was argued at that time that classical mechanics approach could not describe this effect. We have demonstrated using finite-element computer simulations with classical rheological laws that such stress distributions can be found within a classical approach. Furthermore, the found solutions exhibit the same behaviours and depend on the building process [6]. Similar results have been found recently experimentally and are in agreement with simulations based on discrete element method [7].

Thanks to this success, we have decided to pursue this numerical approach to study the avalanche phenomena using a finite element code and classical rheological behaviours in the following way: an isocele triangular pile of granular matter is built on a horizontal rigid base, under some vertical gravity g, *i.e.* $g=10m/s^2$; the pile slopes, $\theta_1=\theta_2=\theta_o$ , are chosen smaller than the friction angle $\varphi$; the stress distribution is then found using finite element technique, with an elastic perfectly plastic constitutive law. Then the evolution of the stress field distribution is followed as the base is submitted to a rotation $\delta\theta$, imposing that one of the pile slopes becomes $\theta_1=\theta_o+\delta\theta$ and the other $\theta_2=\theta_o-\delta\theta$. If we admit that computation diverges when the failure limit is just passed, we are able to determine the maximum angle of repose $\theta_M$.

This paper shows that failure can occur for maximum angle of repose $\theta_M$ much larger than $\varphi$ when a Drücker-Prager law is used. However this result does not hold true when a Mohr-Coulomb yield function is used. The difference between these two behaviours are explained and interpreted as linked to the part plaid by the intermediate principal stress $\sigma_2$: it turns out that in the first case of a Drücker-Prager criterion, a slight variation of $\sigma_2$ around $\sigma_3$, modifies strongly the yield criterion when it is written in the ($\sigma_1$, $\sigma_3$) plane; this does not occur with the Mohr-Coulomb criterion, since it postulates a symmetry breaking as soon as $\sigma_2 \neq \sigma_3$. This analysis is confirmed by the simulations which demonstrate that $\sigma_2$ plays an important part () in the case of the calculations with Drücker-Prager yield function and no important part with the Mohr-Coulomb one.

It is then argued that the spontaneous symmetry breaking of the Mohr-Coulomb criterion is rather unphysical, since it occurs as soon as $\sigma_2 - \sigma_3 \neq 0$. It is then rather more plausible that it exists a small range $\delta$, *i.e.* $-\delta < (\sigma_2 - \sigma_3)/\sigma_3 < \delta << 1$, where the real behaviour does preserve the equivalence between $\sigma_2$ and $\sigma_3$, and for which a Drücker-Prager -like criterion shall be written instead of the Mohr-Coulomb one. The present computations imply in this case that stability of the free-surface slope can be ensured even for slope angles $\theta$ slightly larger than the friction angle $\varphi$, that the difference $\theta - \varphi$ is explained by the role plaid by $\sigma_2$ and that the maximum of difference $\theta_M - \varphi$ is a function of $\delta$.

So, this paper demonstrates that slopes inclined more than $\varphi$ can be metastable most likely. Furthermore, as these simulations are based on classical soil mechanics and as they show that stress field adapt itself spontaneously to preserve pile equilibrium when the pile is rotated, they could explain why instability is reached only when $\theta > \varphi$ when rotating a pile. It means that it may explain the whole avalanche process.

The paper is divided in few parts. In the first part, characteristics of the computer simulation technique is recalled. The second part describe the results from simulations using an elastic-plastic modelling, with either a Drücker-Prager yield function (§-2.1) or a Mohr-Coulomb yield function (§-2.2). Results are discussed in the third part, where the role plaid by the intermediate principal stress $\sigma_2$ is emphasised, with $\sigma_1 < \sigma_2 < \sigma_3$, and $|\sigma_1| > |\sigma_2| > |\sigma_3|$ since s is negative.

## 1. Computer simulation technique

### *1.1 Numerical Models and rheological laws*

In order to study the stability of the triangular soil pile, 3d finite-element simulations have been performed with a continuous-translation symmetry along Oz, i.e. along the horizontal axis perpendicular to the isocele triangle. Previous studies [8] (Pastor and Quecedo 1995) on cases with strain localisation have shown the important role of mesh orientation and mesh size on the obtained results. So we have kept in mind that these parameters affect the quality of the obtained results though in our computations no strain localisation is observed. So, three meshes with approximately the same number of nodes and elements (600 to 700), with different structures have been used (Figure 2). In all cases linear quadrilateral and very few triangular elements have been used. The free surface of the pile makes an angle of 25° with the horizontal. The base of the mesh is fixed.

An isotropic linear elastic-perfectly plastic model with either an associated Drücker-Prager yield function or with a Mohr-Coulomb yield function has been used to model the soil behaviour. In both cases, the friction angle φ was equal to 30°. Though the Drücker-Prager model is not the most realistic one for soils, it can represent some fundamental aspects of granular soil behaviour, as it will be discussed later. Moreover, we have found that the choice of a non associated flow rule does not affect the obtained results.

## 1.2 Initial state and Convergence criteria Procedure of evolution:

As the initial stress state which depends on the history of the pile's construction is one of the key parameters in the response of the soil, the role of this factor has been analysed. Three strategies have been studied:

STR1: Step by step construction of horizontal layers of the pile (only performed for Mesh#1),

STR2: Instantaneous construction of the whole pile,

STR3: Instantaneous construction of the whole pile under a small gravity acceleration followed by the gradual gravity increase.

Similar criteria for stopping the computations have been used for all cases. They are based on the non convergence of unbalanced forces (**F**) and iterative displacements (**u**). Labelling $|\mathbf{A}|=(\sum_i A(i))^{1/2}$, results are accepted at time step n+1 and iteration k+1, as long as:

$$|\mathbf{u}^{k+1}_{n+1} - \mathbf{u}^{k}_{n+1}| / |\mathbf{u}^{0}_{n+1}| < Tol_u \text{ and } |\mathbf{F}^{k+1}_{n+1}| / |\mathbf{F}^{0}_{n+1}| < Tol_F$$

where $Tol_u$ and $Tol_F$ are given positive numbers. Small values of $Tol_u < 10^{-4}$ have always been chosen and it has been observed that this criterion is always verified. So, it has been found that all computations were stopped with respect to the convergence test on the unbalanced forces. $Tol_F$ has been chosen always the same for all computations which have been performed using the same rheological law (Drücker-Prager or Mohr-Coulomb) ; but different values have been chosen for computations with these two different laws, *i.e.* $Tol_F = 10^{-4}$ for the Drücker-Prager yield function and $Tol_F = 10^{-2}$ for the Mohr-Coulomb law. This is due to the fact that calculations converge much easier with the first law than with the second one. The reason of the better convergence of the first calculation will be explained in the next sections.

## 1.3 Procedure of evolution:

Monotonic and cyclic rotations of the pile have been modelled. (As a matter of fact, for the sake of simplicity, the pile has been considered as fix but its rotation has been modelled by rotating the gravity in the other direction). In the case of monotonic rotation, pile was inclined by steps of about 0.5°, till no solution was found. In the cases of cyclic rotation, the pile was rotated about 1° at each time step and two strategies where adopted, *cf.* Fig. 3. In the first one, one or few cycles are performed with a total rotation amplitude of $4° < \varphi - \theta_0$, then rotation is pursued either to the left or to the right ; while in the second strategy the first half cycle rotates $8° > \varphi - \theta_0$ , then changes of sign till breaking. The chronology of the cyclic rotations is given in Fig. 3.

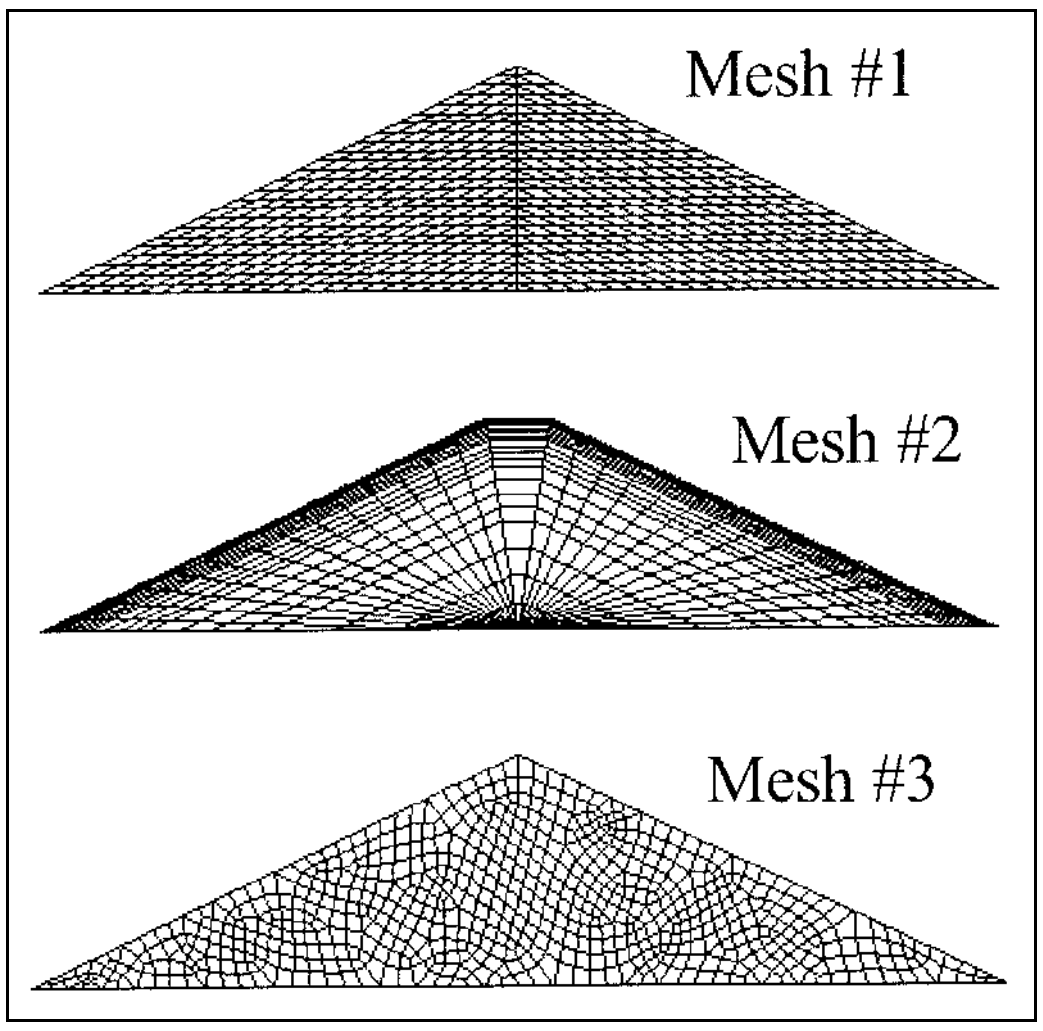

***Figure 2:*** *The three meshes used in the analyses*

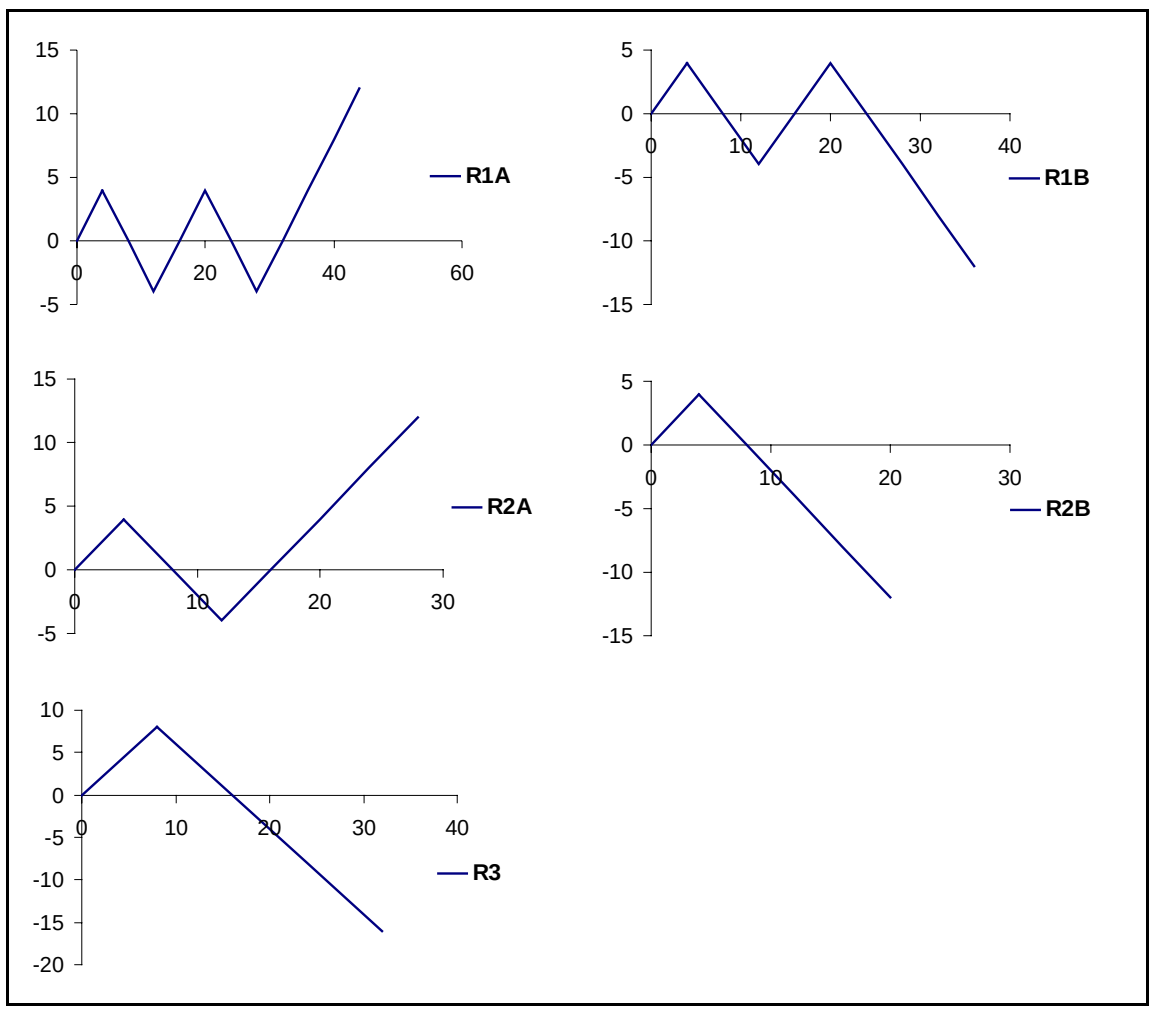

***Figure 3:*** *Chronology of $\delta\theta$ variation for different cyclic rotation tests*

## 2. Computer results

### *2.1 Computer results using Drücker-Prager yield function*

All computations with the Drücker-Prager elastic-plastic model show that, depending on the mesh and the initial condition, non convergence is reached at different values of rotation $\Delta\theta$. The greatest $\Delta\theta$ which has been reached for each computation is given in Table 1. As it can be seen on the same table, non-associated flow rule results always in less admissible rotation.The maximum rotations obtained after a cyclic rotation of the pile for cases Mesh #2-STR2 and Mesh #2-STR3 are given in Table 2. They correspond always to a slope angle $\theta_1$ larger than $\varphi_{DP}$ , *i.e.* $\varphi_{DP}$ =30°.

| Mesh # | $\Delta\theta$ for STR1 | $\Delta\theta$ for STR2 | $\Delta\theta$ for STR3 |
|---|---|---|---|
| 1 | 21°(15.5°) | | |
| 2 | - | 10°(8.5°) | 7.5° |
| 3 | - | | >11° |

***Table 1:*** *Maximum pile rotation $\Delta\theta$ obtained for the monotonic rotation of the pile with an Associated Drücker-Prager yield function. The values between parentheses correspond to a non-associated flow rule resulting in no volumetric plastic strain.*

| | STR2 | STR3 |
|---|---|---|
| R2A | 7° | 7° |
| R2B | 7° | 10° |
| R3 | 5° | >10° |

***Table2:*** *Maximum rotation $\Delta\theta$ obtained for the Mesh #2 after cyclic rotation with a Drücker-Prager yield function.*

In Figures 4 and 5, the elastic (grey) and plastic (black) zones for cases Mesh #1-STR1 and Mesh #2-STR3 are given. As it can be noticed, in case Mesh #1-STR1, the initial state, just after the construction step, includes both elastic and plastic zones. In this case, the computation can be continued until the whole pile is plastified, while in case Mesh #3-STR3 the computation diverges as soon as an important plastic zone is generated on one side of the pile.

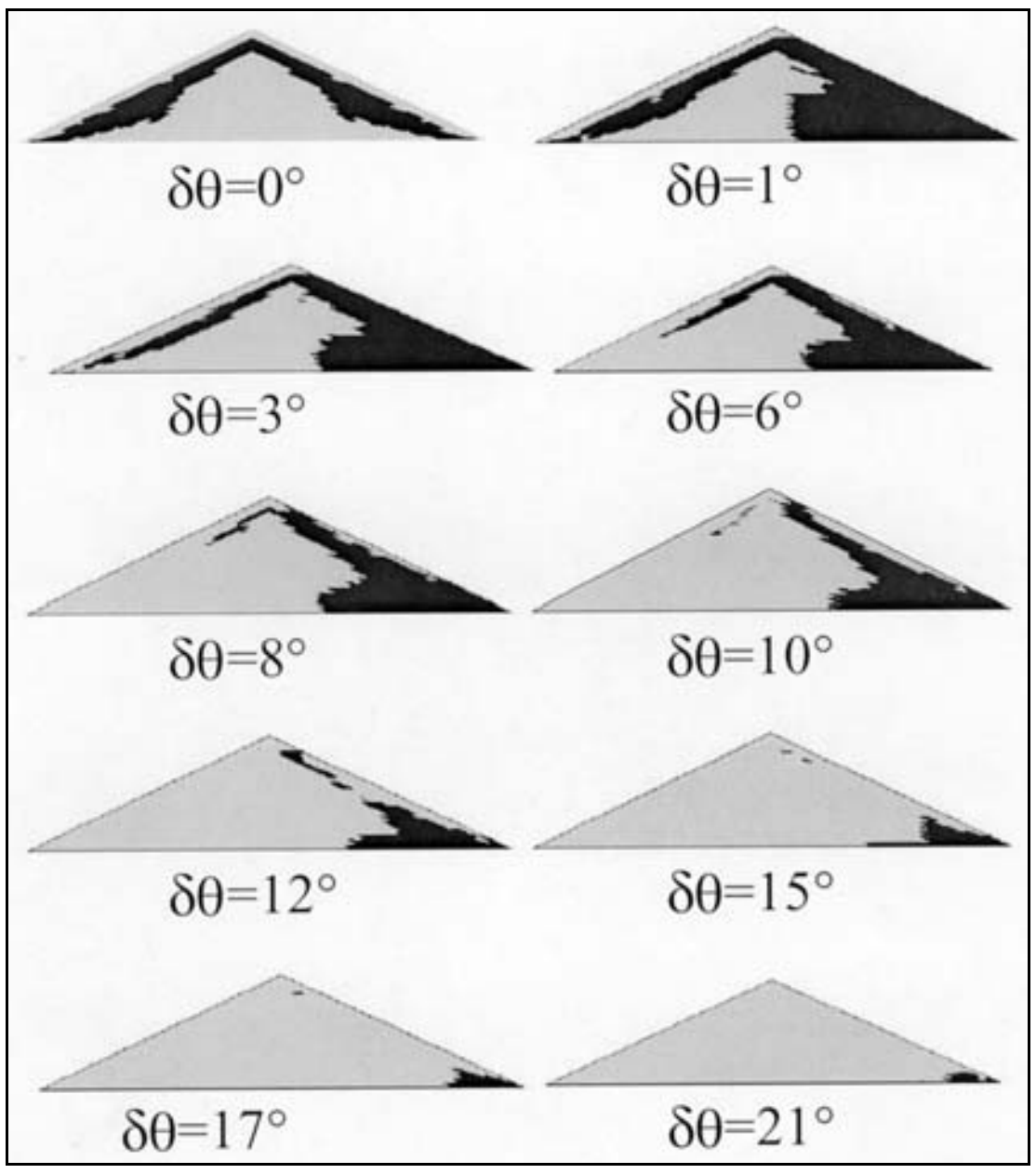

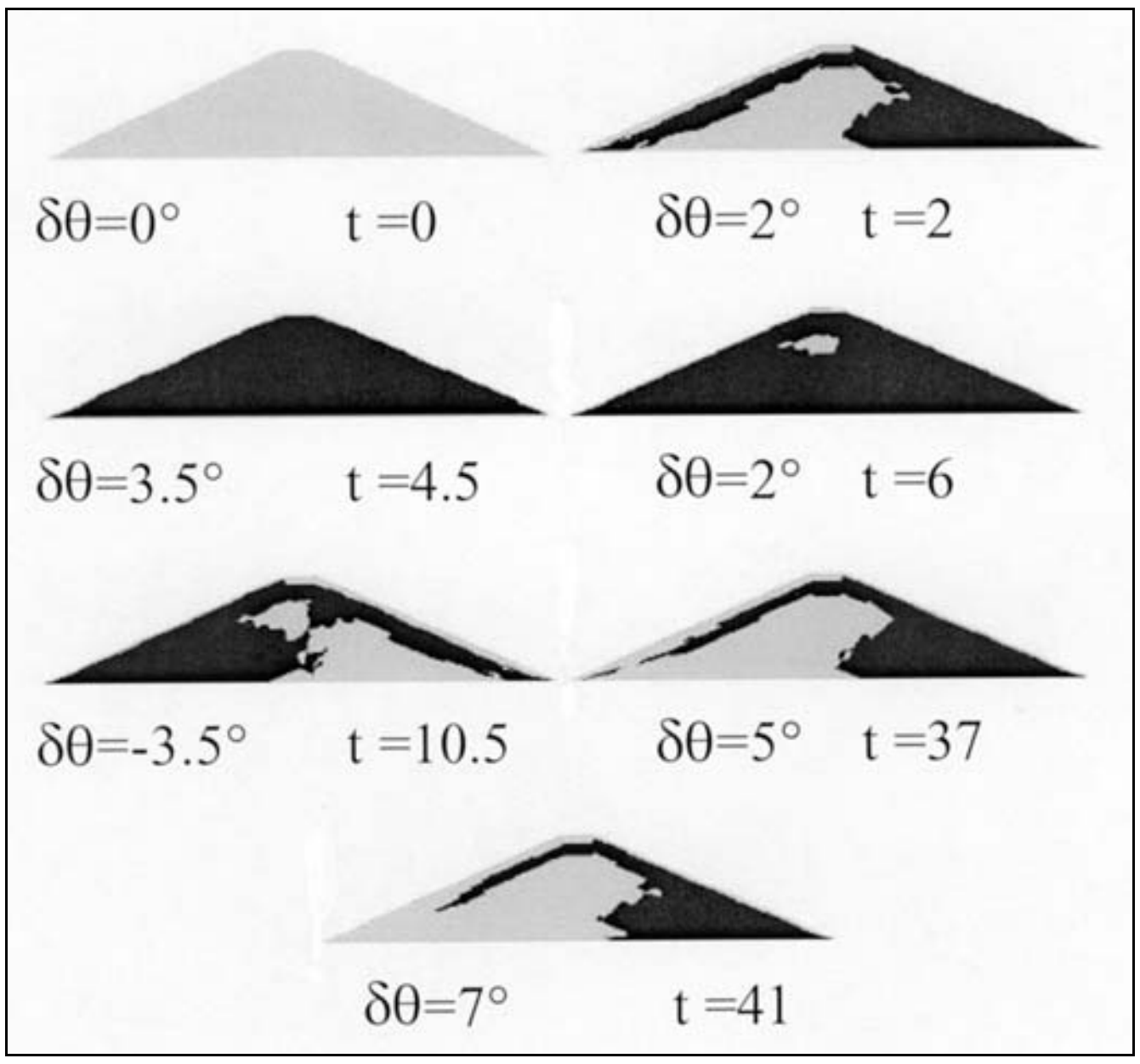

***Figure 4:*** *Distribution of elastic (black) and plastic (grey) zones during monotonic rotation of the* Mesh#1-STR1 *case.*

***Figure 5:*** *Distribution of elastic (black) and plastic (grey) zones during the cyclic rotation* (R1A) *of case* Mesh#2 STR2. *The parameter t describes the chronology of the rotation. This simulation corresponds also to Fig. 6.*

Figs 6-8 report the evolution of the stress- and/or the strain- distributions as a function of the angle the pile has rotated. For instance, Fig. 6 has been obtained by imposing a first cycle of rotation with amplitude $\Delta\theta=4°$ , i.e. $\theta=25°\rightarrow29°\rightarrow21°\rightarrow29°$, then the rotation has been increased till non convergence, i.e. till 33°. The graphs of Fig.6 correspond to the following set of inclinations $26°\rightarrow28° \rightarrow28°\rightarrow 26°\rightarrow24° \rightarrow22°(\rightarrow21°)\rightarrow22° \rightarrow24°\rightarrow26°\rightarrow28° (\rightarrow29°) \rightarrow28°\rightarrow26° \rightarrow24°\rightarrow22° (\rightarrow21°) \rightarrow22°\rightarrow24°\rightarrow26°\rightarrow28°\rightarrow30°\rightarrow32°$, where the minimum and maximum values have been indicated in between parentheses. The calculation did not converge for $\theta=33°$. The same rotation history has been used for Fig. 5. In Fig. 5, the elastic *vs.* plastic zones are reported only, whereas the components $\sigma_{xx}$ and $\sigma_{yy}$ of the stress field are also reported in Fig. 6, (x being the horizontal, y being the vertical). It is worth noting in Fig.6 that the material of the pile remains fully elastic for intermediate inclination, *i.e.* $22°<\theta<28°$, after the first cycle has been performed. One notes also that the stress fields, $\sigma_{xx}$ and $\sigma_{yy}$, do not evolve significantly during the rotation. Indeed, we will show in the following section, *i.e.* section 3, that the main way the system chooses to adapt itself to the inclination change is by changing its $\sigma_{zz}$ stress component, since this is possible in the case of the Drücker-Prager law.

Fig. 7 gives an example of the stress evolution during a continuous increase of inclination from 25° to 31.5°, with the Drücker-Prager law. The systems did not converge for $\theta=32.5°$. Once more, one does not observe also any significant change either in the distribution of the stress components $\sigma_{xx}$, $\sigma_{xy}$, $\sigma_{yy}$, or in the distribution of the plastic/elastic zones, when passing over $\theta=\varphi$.

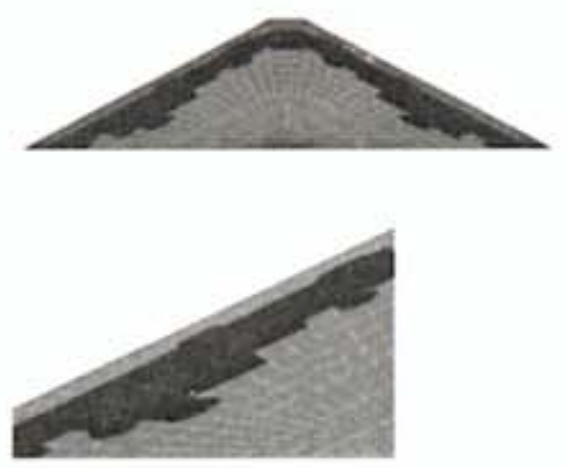

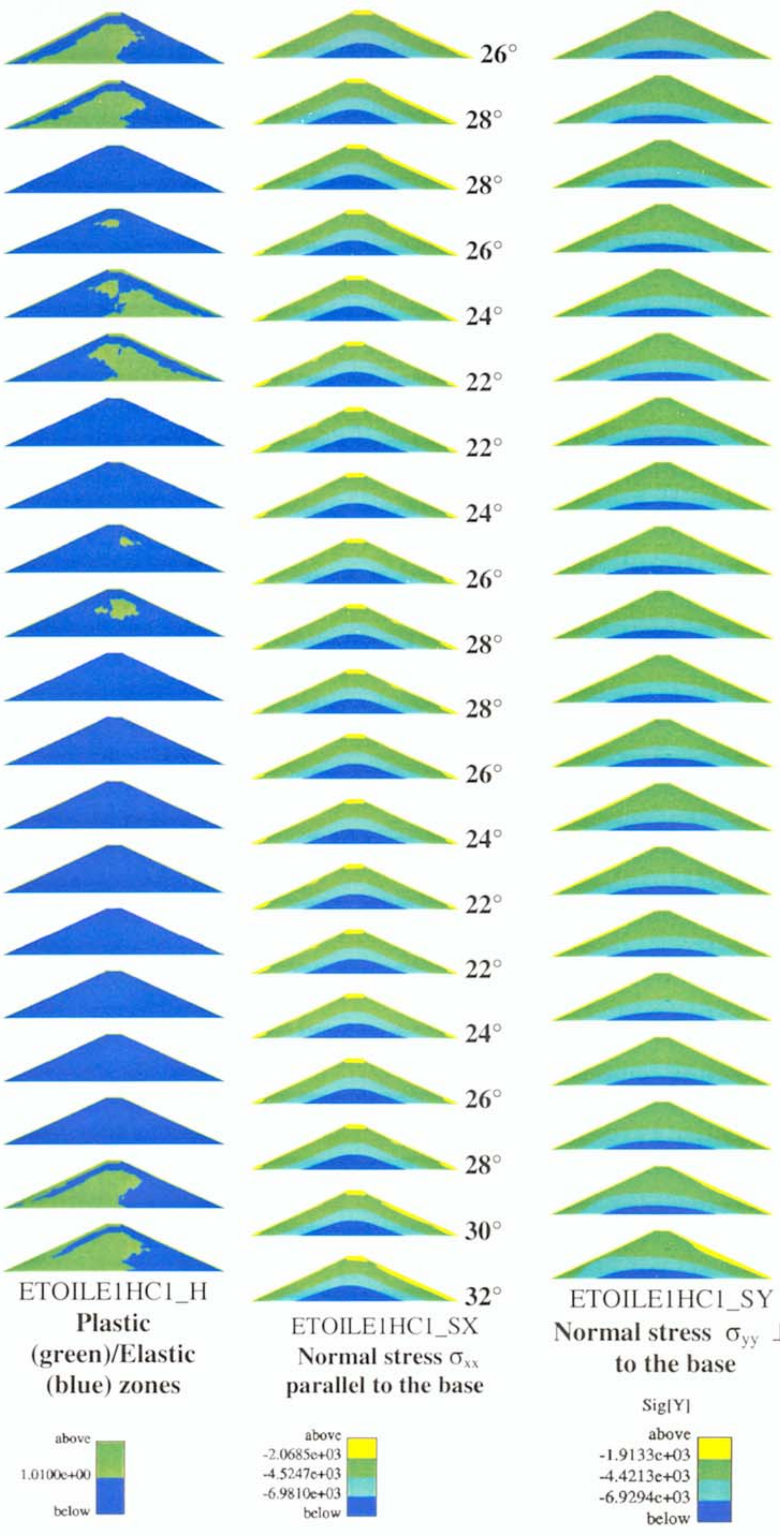

***Figure 6:*** *Each column corresponds to the evolution of the distribution of a peculiar stress component in a 2d triangular pile as a function of the rotation angle.*

*The steps correspond to slope angles:* 26°→28°(→29°)→28°→26°→24°→22°→(21°)→22°→24°→26°→28°→(29°)→28°→26°→24°→27°→28°→29°→30°→32°

***Parameters:*** *Drücker-Prager law. Friction angle* $\varphi$=30°; *Starting slope* $\theta_o$= 25°; $\theta_{max} \approx 32°$ ; $\delta\theta \approx 2°$ *at each rotation step.*

***Column 1:*** *the mesh and the resolution*
***Column 2:*** *plastic zone*
***Column 3:*** *distribution of stress parallel to the base of the pile*
***Column 4:*** *distribution of stress perpendicular to the base of the pile.*

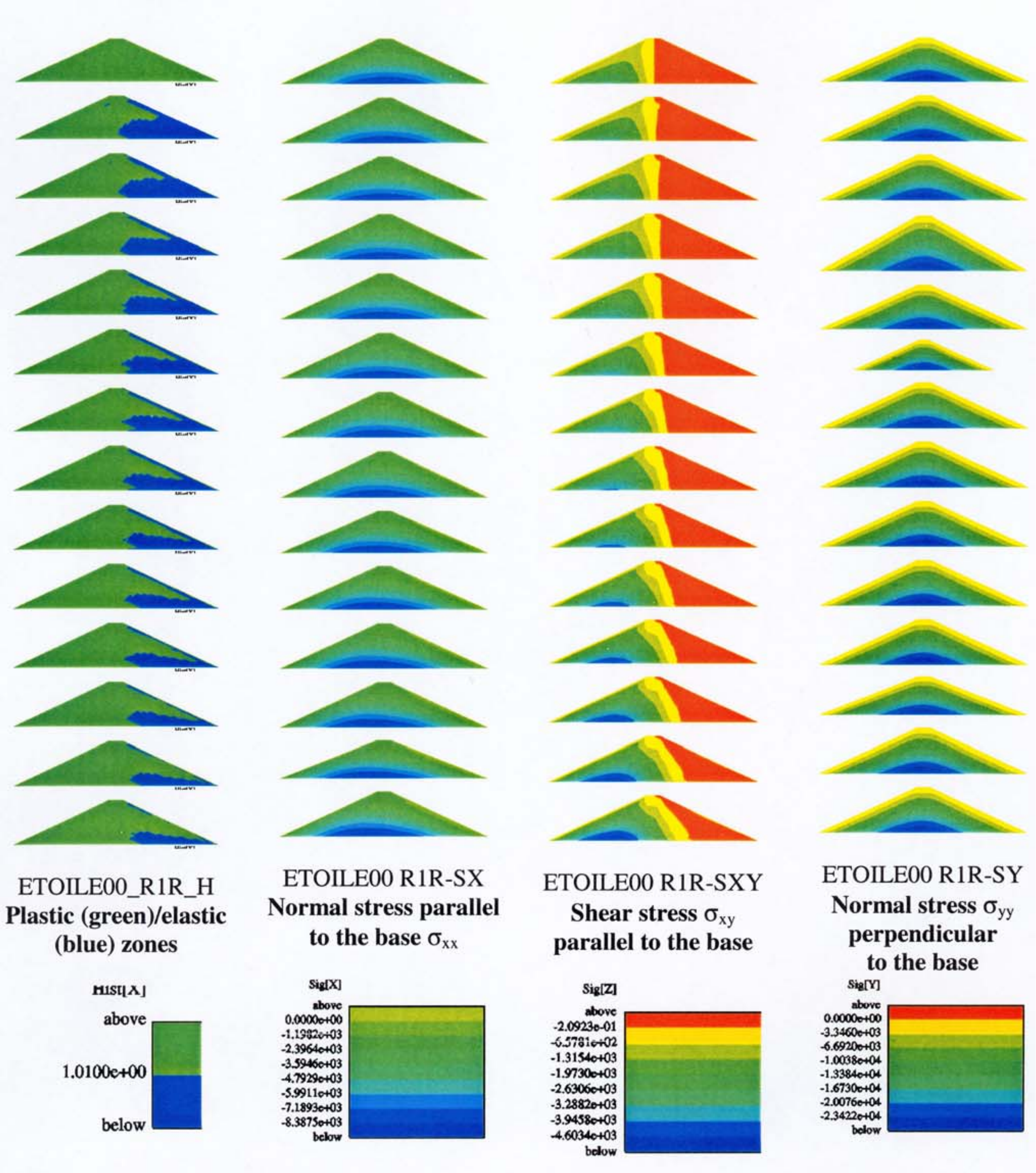

***Figure 7: Evolution of stress distribution as a function of the rotation. Friction angle** φ=30°. **Drücker-Prager law, Mesh #2.***

*Column1: evolution of the plastic zone;*

*column 2: evolution of contours of the normal stress on planes perpendicular to the base* $\sigma_{xx}$ *;*

*column 3: evolution of contours of the shear stress on planes parallel to the base* $\sigma_{xy}$ *;*

*column 4: evolution of contours of the normal stress on planes parallel to the base* $\sigma_{yy}$ *;*

*Parameters: starting slope angle:* $\theta_o$= 25° *on both sides ; Drücker-Prager law with* φ=30°*;* *maximum slope angle:* $\theta_o$= 32°; *rotation of* δθ *per figure:* δθ≈0.5° *at each step*

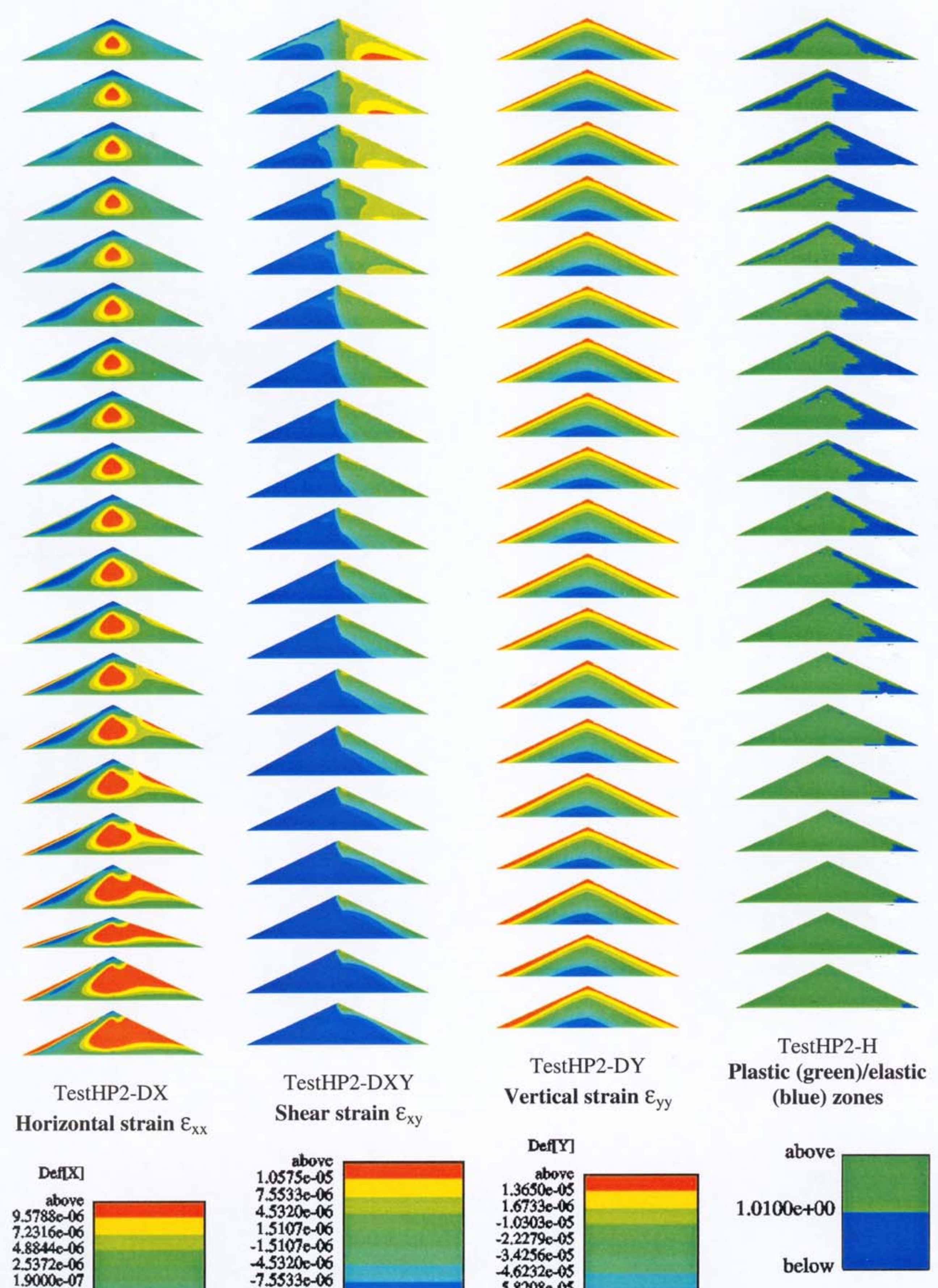

***Figure 8**: columns 1-4. For captions see next page*

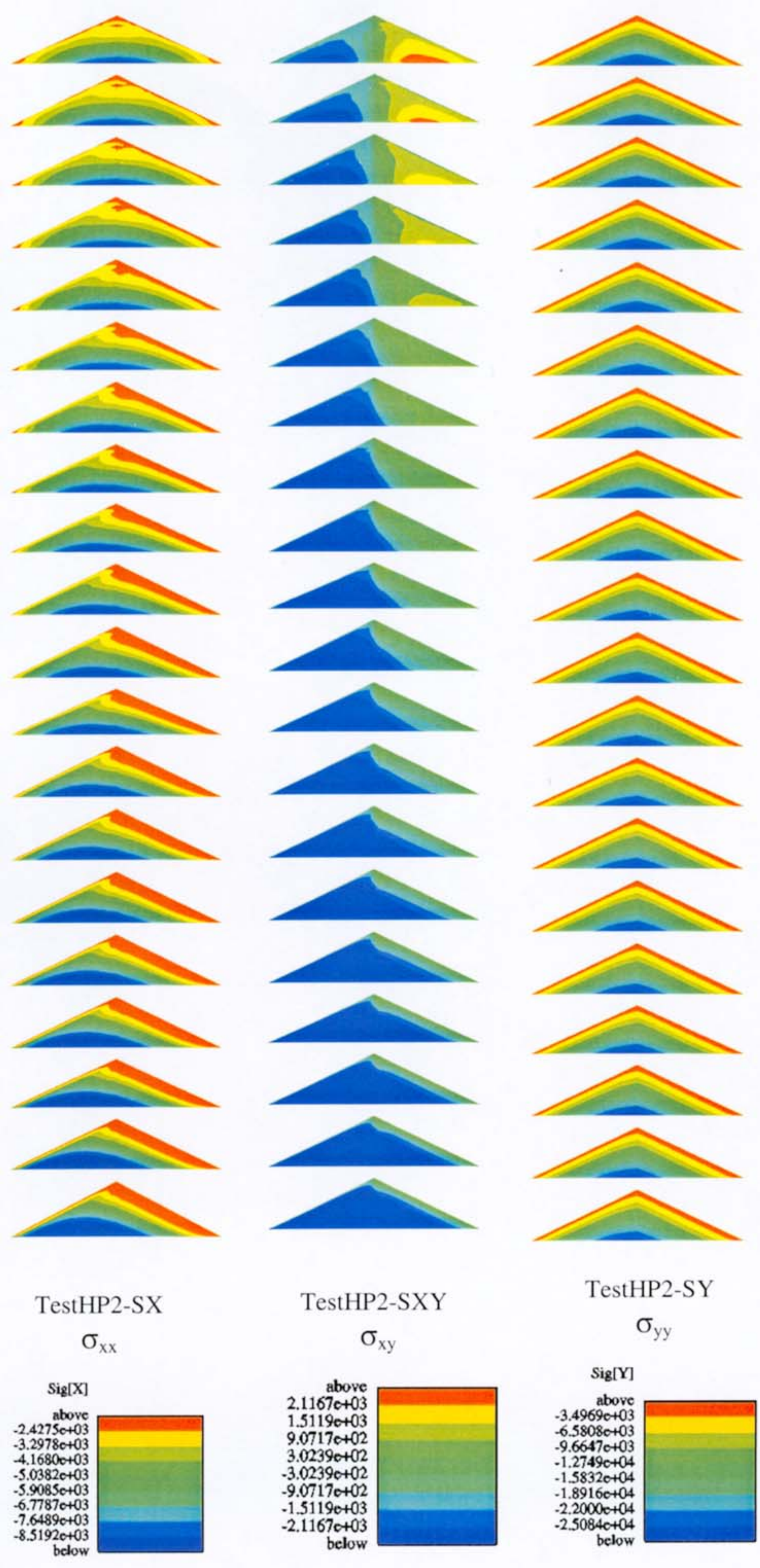

***Figure 8:*** *Computed evolution of the stress and strain distributions in a 2d symmetric triangular pile as a function of rotation; Drücker-Prager law; friction angle* $\varphi$=30°; *initial slopes:* $\theta_o$=±25°; *ending slopes:* $\theta_{max}$= 48° *and* $\theta_{min}$=-2°.
$\delta\theta\approx$1° *between two vertical figures.*
***Column 1:*** *evolution of* $\varepsilon_{xx}$ *(the normal strain on planes perpendicular to the base) contours.*
***Column 2:*** *evolution of* $\varepsilon_{xy}$ *(shear strain on planes parallel to the base) contours.*
***Column 3:*** *evolution of* $\varepsilon_{yy}$ *(normal strain on planes parallel to the base) contours.*
***Column 4:*** *evolution of the elastic/plastic zones.*
***Column 5:*** *evolution of* $\sigma_{xx}$ *contours ( normal stress on planes perpen-dicular to the base)*
***Column 6:*** *evolution of* $\sigma_{xy}$ *contours (shear stress on a planes parallel to the base).*
***Column 7:*** *evolution of* $\sigma_{yy}$ *contours (normal stress on planes parallel to the base).*

Fig. 8 reports the complete evolution in the (x,y) plane of the stress ($\sigma_{xx}$, $\sigma_{xy}$, $\sigma_{yy}$) and the strain ($\varepsilon_{xx}$, $\varepsilon_{xy}$, $\varepsilon_{yy}$,) for a case where the evolution could be studied far above the friction angle since the maximum slope angle which was achieved was $\theta_{max}$=25+21°=46° ; as in Fig. 7, no drastic evolution is found; one remarks the continuous decrease of the size of the elastic zone when the pile is rotated. One can also remark that the plastic zone spans over the whole pile at large inclination; the pile breaks finally when the elastic zone vanishes.

## *2.2 Computer results using Mohr-Coulomb yield function*

It is worth noting at first that all computer calculations performed with the Mohr-Coulomb yield function converge less rapidly than the ones with the Drücker-Prager criterion and the results are obtained with less accuracy. A second interesting point is that no solution has ever been found for piles with a slope angle $\theta$ larger than $\varphi$ , and convergence stops for $\theta=\varphi$-0.5°. Two typical examples of the evolution of the stress distribution in the (x,y) plane are given in Figs. 9 & 10; they have been obtained with the two different laws, *i.e.* Mohr-Coulomb and Drücker-Prager. They do not exhibit any great difference, nor they differ significantly from the results of the preceding subsection. However, we will see in the next section that Fig. 10 allows to understand why the free surface of the pile can be more inclined than $\varphi$ in the case of the Drücker-Prager law.

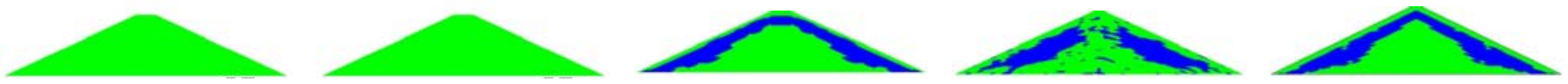

| Mesh #2 - STR3 Mohr-Coulomb | Mesh #2 - STR2 Mohr-Coulomb | Mesh #2 - STR2 Drücker-Prager | Mesh #1 - STR1 Mohr-Coulomb | Mesh #1 - STR1 Drücker-Prager |
|---|---|---|---|---|

***Figure 9:*** *The distribution of the plastic zones depends on the construction history and the constitutive law, results obtained at the end of the construction phase*

We report in Fig. 11 a typical evolution of the deformation of the pile during a continuous rotation; there is not much difference for the two different laws, i.e. Drücker-Prager and Mohr-Coulomb.

## 3. Discussion

As similar results have been obtained with different mesh refinements and/or mesh structures, the results are supposed to be representative and show that metastable piles can be generated with an angle of repose $\theta_M$ larger than the friction angle $\varphi$ in the case of a Drücker-Prager yield function. However these results do no longer hold true when the pile is assumed to obey a Mohr-Coulomb criterion for which we have obtained always $\theta_M \approx \varphi$ : in this case, the slope never overpasses the friction angle $\varphi$. This point will be discussed and explained below; but let us make first the two following remarks:

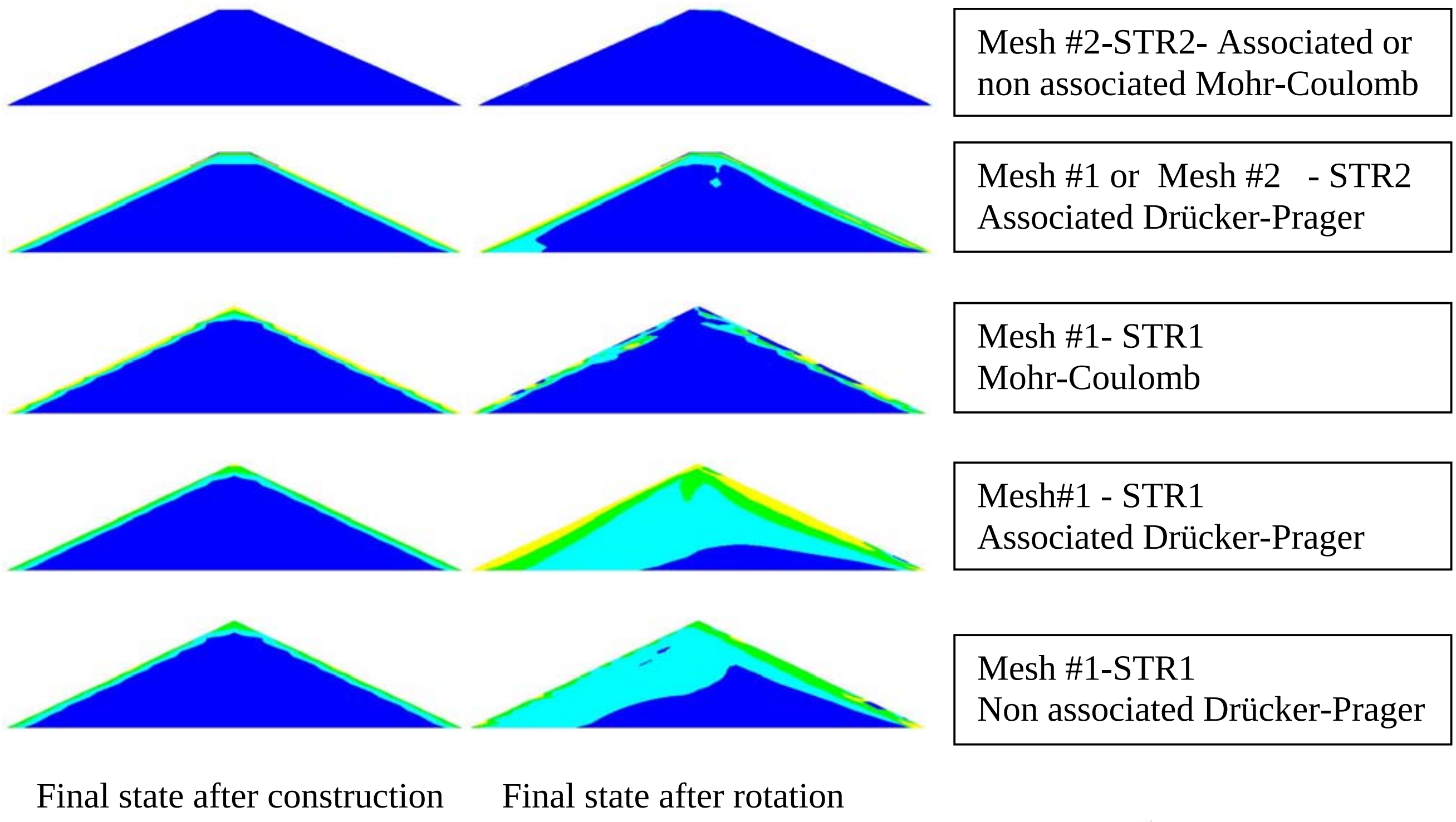

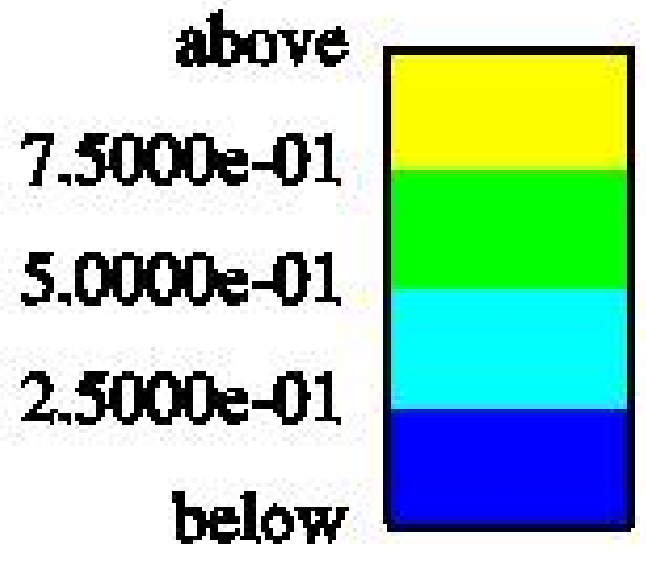

***Figure 10:*** *The distribution of* b=($\sigma_1$-$\sigma_2$)/($\sigma_1$-$\sigma_3$) *depends on the construction history, of the constitutive law and of the rotation; the reported results have been obtained at the end of the construction (left) and at the maximum rotation (right).*

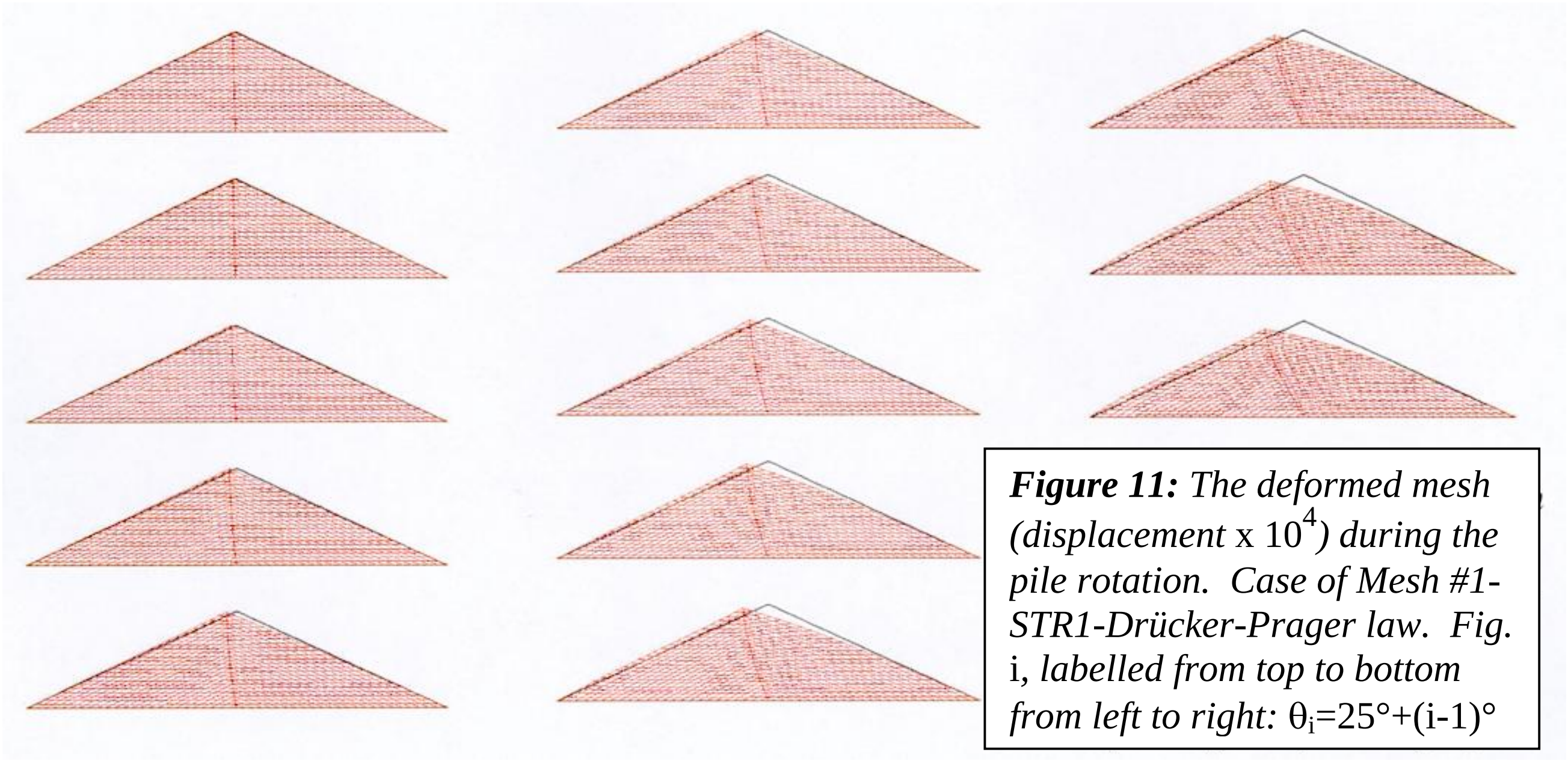

***Figure 11:*** *The deformed mesh (displacement* x $10^4$*) during the pile rotation. Case of Mesh #1-STR1-Drücker-Prager law. Fig.* i, *labelled from top to bottom from left to right:* $\theta_i$=25°+(i-1)°

- It is worth noting that, for all calculations, the plastic limit is always achieved near the free surface; this is in agreement with the work of Cantelaube & Goddard [9], which has shown that elastic solution is not possible near the inclined free surface.

- The initial plastic zone in the pile centre is linked to the building process; this part becomes elastic as soon as the pile has undergone a cyclic rotation. It is only when rotation becomes larger than $\varphi$-$\theta_o$ that a central plastic zone reappears in the case of the Drücker-Prager law. For the Mohr-Coulomb law similar remarks can be made.

### *3.1 Difference between the Mohr-Coulomb and the Drücker-Prager yield functions*

The Mohr-Coulomb criterion takes never into account the intermediate principal stress $\sigma_2$, so that yielding problem remains always a planar problem; this imposes that the direction of the intermediate stress plays a specific role and has to be identified prior to apply the yielding criterion. Indeed, in the Mohr-Coulomb approach, yielding occurs if the following condition is satisfied:

$$(\sigma_1-\sigma_3)/(\sigma_1+\sigma_3)=\sin(\varphi_c) \qquad (1)$$

without referring to $\sigma_2$, and where $\varphi_c$ is the friction angle of the Mohr-Coulomb criterion. One knows that $\varphi_c$ shall be identified to the friction angle of the critical state, in the approach of Schofield-Wroth-Roscoe.

On the contrary, in the Drücker-Prager approach, the three principal stresses and stress directions play an equivalent role since the yielding criterion writes:

$$(\sigma_1-\sigma_2)^2+ (\sigma_1-\sigma_3)^2+ (\sigma_2-\sigma_3)^2 = \alpha^2 (\sigma_1+\sigma_2+\sigma_3)^2 \qquad (2)$$

where $\alpha$ is the Drücker-Prager parameter.

In order to make the two descriptions as equivalent as possible, and to define a friction angle, one can consider the case when the stress ratio $s=\sigma_2/\sigma_3$ is fixed and identify the two criteria. For instance, if one considers the case $\sigma_2=\sigma_3$ , equating the two criteria leads to the condition:

$$\sin(\varphi_{s=1}) = 3\alpha_1 / (2\sqrt{2}+\alpha_1) \qquad (3.a)$$

However one could have chosen to identify the two criteria when $\sigma_2=(\sigma_1+\sigma_3)/2$; or when $\sigma_2=\sigma_1$ ; in these cases, identification of the two criteria leads respectively to the two equations:

$$\sin\varphi_{s'} =\alpha_2 (3/2)^{1/2} \qquad (3.b)$$

$$\sin\varphi_{s''} =3\alpha_3/(2\sqrt{2}-\alpha_3) \qquad (3.c)$$

One can generalise this approach and can conclude that the identification of the two criteria is always possible for a given $s=\sigma_2/\sigma_3$ , but that the relation between $\varphi_s$ and $\alpha$ depends on $s=\sigma_2/\sigma_3$ [10]. So, once the identification is performed for a given $s_o=[\sigma_2/\sigma_3]_o$ , the two yield criteria will differ when the ratio $\sigma_2/\sigma_3$ is varied, *i.e.* for a different ratio $s_1=[\sigma_2/\sigma_3]_1$ . It means that the maximum slope of a pile which obeys the Drücker-Prager criterion will depend on $\sigma_2/\sigma_3$. In order to make the problem more evident, let us restate it as follows: let us consider that the material we are studying obeys truly the Mohr-Coulomb approach, but let us also consider that we have chosen to describe it with the Drücker-Prager approach. In this case the system is described by a unique $\varphi$, but we want to describe it with Eq. (2); this may be done, but this requires that we chose an $\alpha$ which varies with the stress ratio $s=\sigma_2/\sigma_3$.

Conversely, let us consider that the material we are studying obeys truly the Drücker-Prager law, but let us also consider that we have chosen to describe it with the Mohr-Coulomb approach. In this case the system is truly described by a unique $\alpha$, but we want to describe it with Eq. (1); this may be done, but this requires that we chose a $\varphi$ which varies with the stress ratio $\sigma_2/\sigma_3$. So, if we apply now this result to the stability of a slope of a material which obeys the Drücker-Prager criterion, the preceding result imposes that the maximum slope will depend on the ratio $\sigma_2/\sigma_3$.

One can evaluate the range of variation of the maximum slope in the case when the material obeys the Drücker-Prager yield function by choosing a given $\alpha$ ; then applying Eqs. (3.a), (3.b), & (3.c) allows to find the maximum slope $\varphi_s$ for the three different values of $\sigma_2$, *i.e.* $\sigma_2=\sigma_3$, $\sigma_2=(\sigma_1+\sigma_3)/2$ & $\sigma_2=\sigma_1$. They are reported in Table 3 for different values of $\alpha$. Fig. 12 reports also the variation of the maximum slope $\theta_M$ as a function of $s=\sigma_2/\sigma_3$ (Fig. 12.a) or as a function of $\theta_{Ms}$=Arcsin[(s-1)/(s+1)] (Fig. 12.b).

| | b | $\alpha$=0.46378 | $\alpha$=0.56569 | $\alpha$=0.66861 | $\alpha$=0.77128 | $\alpha$=0.87226 |
|---|---|---|---|---|---|---|
| $\sigma_2=\sigma_3$ | 0 | $\varphi_{s=1}$=25° | $\varphi_{s=1}$=30° | $\varphi_{s=1}$=35° | $\varphi_{s=1}$=40° | $\varphi_{s=1}$=45° |
| $\sigma_2=(\sigma_1+\sigma_3)/2$ | 0.5 | $\varphi_s$ =34° 37' | $\varphi_s$ = 43° 51' | $\varphi_s$ = 54° 48' | $\varphi_s$ = 70° 50' | No solution >90° |
| $\sigma_2=\sigma_1$ | 1 | $\varphi_s$ = 36° 03' | $\varphi_s$ = 48° 36' | $\varphi_s$ = 68° 14' | No solution >90° | No solution >90° |
| $\delta\varphi/(\delta\sigma_2/\sigma_3)$ from Eq. 4.b | | 13° | 12.4° | 11.7° | 11.0° | 10.1° |
| $\delta\varphi/(\delta\sigma_2/\sigma_3)$ from Fig. 12.b | | 12.9° | 12.3° | 11.7° | 11.0° | 10.1° |

***Table 3:*** *maximum slope $\varphi$ of the free surface of a medium obeying the Drücker-Prager yield function for different values of $\alpha$ and for different values of the intermediate stress field* $\sigma_2$, *i.e.* $\sigma_2=\sigma_3$, $\sigma_2=(\sigma_1+\sigma_3)/2$, $\sigma_2=\sigma_1$. *The 2 last lines corresponds to the variation of the friction angle when starting from $\sigma_2=\sigma_3$ and increasing slightly $\sigma_2$ and not $\sigma_3$ , for the different $\alpha$ or $\varphi_{s=1}$; these values are numerical estimates from either* Eq. (4.b) *or from simulations; they mean that a slope variation of* 3° *about correspond to* $\delta\sigma_2/\sigma_3$=0.3 .

To enlighten the real behaviour of the material which obeys the Drücker-Prager law, it is worth considering the problem of a system which starts with the condition $\sigma_2=\sigma_3<\sigma_1$ and which is at the limit of stability in the Drücker-Prager meaning. So, labelling $x=\sigma_1/\sigma_3$, this imposes: $2(x-1)^2=\alpha^2(x+2)^2$ ; or $(x-1)/(x+2)=\alpha/\notin 2$. Let us now perturb this system and increase slightly $\sigma_2$ so that $\sigma_2$ becomes $\sigma'_2=\sigma_2+\delta\sigma_2$, keeping $\sigma_3$ constant . This allows $\sigma_1$ to become larger; it can reach now the value $\sigma'_1=\sigma_1+\delta\sigma_1$ ; $\delta\sigma_1$ can be evaluated by derivation of Eq. (2); one gets then $(x-1)(2\delta\sigma_1-\delta\sigma_2)=\alpha^2(x+2)(\delta\sigma_1+\delta\sigma_2)$. But, as $\alpha^2(x+2)^2=2(x+1)^2$, one gets $(x+2)(2\delta\sigma_1-\delta\sigma_2)=2(x-1)(\delta\sigma_1+\delta\sigma_2)$, or $\delta\sigma_1/\delta\sigma_2=x/2$; in this equation, $\delta\sigma_1$ and $\delta\sigma_2$ correspond to the new yielding condition, whereas x corresponds to the previous yielding one. Labelling $\varphi_{s=1}$ the friction angle which corresponds to the initial condition, *i.e.* $\sigma_2=\sigma_3$ , so $\varphi_{s=1}$ is given by Eq. (3.a) and it is such as $\sin(\varphi_{s=1})=(x-1)/(x+1)$, one finds that:

$$\delta\sigma_1/\delta\sigma_2 = [1+\sin(\varphi_{s=1})] / [2-2\sin(\varphi_{s=1})] \qquad (4.a)$$

which leads to the variation of the maximum slope angle:

$$\delta\varphi=[\cos(\varphi_{s=1})/4] \;\; \delta\sigma_2/\sigma_3 \qquad (4.b)$$

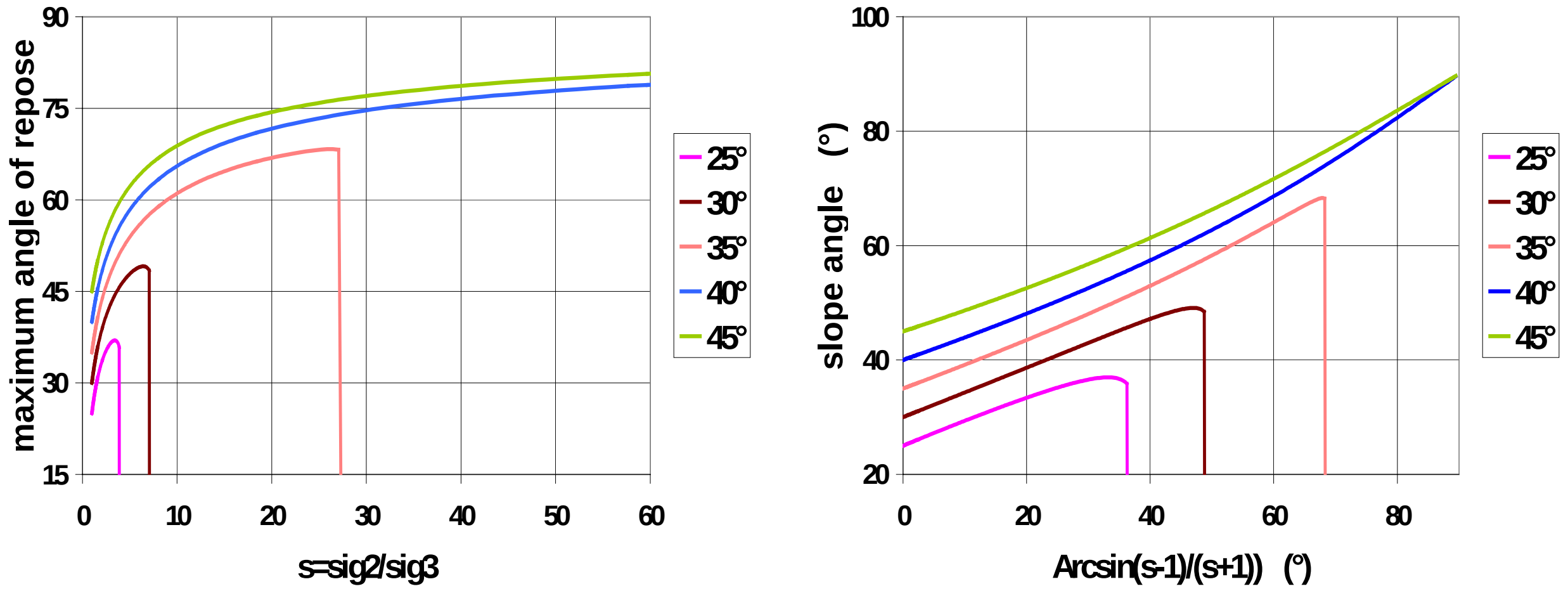

**Figure 12:** *Variations of the maximum slope of a Drücker-Prager material vs.* $s=\sigma_2/\sigma_3$ *(left Figure) or vs.* Arcsin[(s-1)/(s+1)] *(right Figure); all angles are in °. The 5 curves correspond to 5 different values of* $\alpha$ *or* $\varphi_{s=1}$, *i.e.* $\varphi_{s=1}$= 25°, 30°, 35°, 40°, 45°. *Relation between* $\alpha$ *and* $\varphi_{s=1}$ *is given by* Eq. (3.a) .

Estimates, in degree unit, of $\sigma_3\ \delta\varphi\ /\delta\sigma_2$ are reported in the last line of Table 3. To get the precise variation of the angle of repose $\delta\varphi$ which corresponds to a given $\delta\sigma_2/\sigma_3$, one has to multiply these numbers reported in the table by the real value of $\delta\sigma_2/\sigma_3$. From these estimates, it appears that a slight change of $\delta\sigma_2/\sigma_3$ allows a large change of the maximum angle of repose since 30% change of $\sigma_2$ results in a change of $\varphi$ of more than 3 degrees. This allows to understand (i) why calculations using Drücker-Prager law converge rather quickly in general, since convergence can be improved by varying $\sigma_2$ instead of $\sigma_1$ and/or $\sigma_3$ , (ii) why the maximum slope can overpass the "friction" angle of the Drücker-Prager law, which corresponds to $\sigma_2=\sigma_3$.

It is worth noting that $\sigma_1$ can become infinite for any finite $\sigma_3$ if the medium obeys the Drücker-Prager criterion when the ratio $\sigma_2/\sigma_3$ is large enough and as far as $\alpha>1/\not\equiv 2$ , *cf.* Fig 12; this threshold, *i.e.* $\alpha=1/\not\equiv 2$, corresponds to $\varphi_{s=1}>36°\ 87'$. However, as it is demonstrated by Fig. 12.b, the ratio $\sigma_1/\sigma_3$ can only become infinite if $\sigma_2/\sigma_3$ becomes also infinite.

Does this mean that the Drücker-Prager modelling shall be abandoned? We do not believe so. Of course, owing to the present study, the Mohr-Coulomb criterion appears to be much more efficient and more realistic to understand and describe the slope stability of an earth embankment. Conversely, the Drücker-Prager criterion appears to be much more sensitive to the stress field, so that it shall lead to rather unpredictable or uncertain behaviour. However we will argue in two subsections, *i.e.* §-3.3, that the Mohr-Coulomb criterion contains some non physical constrains because it supposes the existence of some spontaneous symmetry breaking as soon as the two minor stresses are only slightly different.

### *3.2 Proof that $\sigma_2$ variations play an important part in the process of finding the stress solution in a pile with $\theta>\varphi$ and obeying the Drücker-Prager law*

Anyhow, we can now come back to Fig. 10. Indeed, one notes that in the case of Drücker-Prager law, the ratio $b=(\sigma_3-\sigma_2)/(\sigma_3-\sigma_1)$ at the end of construction is only slightly greater than the b value in the case of Mohr-Coulomb law, and only at the surface. In addition, pile rotation does not make the ratio b evolve much when the material obeys the Mohr-Coulomb criterion though b decreases a little.

This is not always the case for the Drücker-Prager law for which the evolution of b depends both on the stress state and on the flow rule. As in incremental plasticity, the stress variation is related to the strain increment, the flow rule plays an important role. In our numerical simulations, the plane strain hypothesis is assumed resulting in evolutions of the out of plane stress which are different depending on the dilatancy rule. These two aspects are well illustrated in Fig. 10. For a given mesh and a given construction history (Mesh#1-STR1), b increases more for the non associated flow rule resulting in a greater friction angle $\varphi_s$ which will permit higher rotation of the pile (21>15.5°). This is the proof of the truthfulness of the previous-subsection analysis: the evolution of b does contribute efficiently to the adaptivity of the pile to the new situation when the material obeys the Drücker-Prager law. This is indeed not true for a Mohr-Coulomb material (Fig. 10).

However, one shall note also that the failure angle is not constant for a given Drücker-Prager law, but depends on the pile history too. This means that the pile adaptivity is not maximum but varies with the experimental conditions. It may then vary from one experiment to another one, depending on the boundary conditions. For a given mesh and flow rule the different initial stress states obtained with two construction strategies (Mesh#1-STR1 et STR2) will result in two different evolutions of b. In all cases, the system fails at the rotation corresponding to the friction $\varphi_s$ .

So, rotation of an inclined surface is not a destabilising procedure quite often. However, one cannot find the slope failing at the largest possible value, which would have been here about 50°, for the friction angle which has been used, *i.e.* $\varphi_{DP}$=30°. But the maximum rotations we find correspond well to the computed b values . This explains the values obtained for Mesh #1-STR1 reported in Table 1 ($\varphi_s$ =25+21 corresponds to 1<b<0.75 and ($\varphi_s$ =25+15 corresponds to 0<b<0.25).

### *3.3 Is a breaking of symmetry at $\sigma_2=\sigma_3$ physical?*

Indeed, avalanching experiments, such as those reported in Fig. 1, show that avalanching flow is always observed when a pile is rotated; it means then that the maximum slope $\theta_M$ is always larger than the stopping angle $\theta_s$, leading to $\delta\theta=\theta_M-\theta_s>0$ . A question arises then: Why is $\delta\theta$ never equal to 0?

A part of this answer is probably related to our finding with the Drücker-Prager model: when $\sigma_2=\sigma_3$, there is no reason why the yield criterion of the system should write as the Mohr-Coulomb criterion. On the contrary, it is much more probable that it has to be written as the Drücker-Prager criterion. But, this is no more true as soon as $\sigma_2$ becomes quite different from $\sigma_3$, for which case the Mohr-Coulomb criterion shall be better. This let think that there shall be a small range of $\sigma_2/\sigma_3$ for which $\varphi$ shall vary with $\sigma_2$, so that $\varphi$ is not fix, but varies within a range $\delta\varphi$.

Indeed, it is worth noting that the variation of $\varphi$ with $\sigma_2$ has been already studied experimentally by some authors [11] and that the amplitude $\delta\varphi$ of the variation of $\varphi$ with $\sigma_2$ is $\delta\varphi$=3-5° about which corresponds to $\delta\varphi/\varphi$=10-15%. This fixes the $\sigma_2/\sigma_3$ range for which the variation of $\sigma_2$ is important before the Mohr-Coulomb yield criterion is fully valid.

So, these computations explain, at least partly, results of Figure 1, since they predict that slopes inclined more than $\varphi$ can be meta-stable; and that the meta-stability is not destroyed by rotating the pile. We can then expect that failure occurs for $\theta_M>\varphi$ , and that the avalanche size is $\delta\theta=\theta_M-\varphi$ if we assume that the dynamic friction angle $\varphi_{dyn}$ is equal to the quasi static one $\varphi$. Indeed Figure 1 leads to $\varphi_{dyn}$=32° for Hostun sand; it is just the value obtained with triaxial tests for the quasi static value $\varphi$=32°; so $\varphi_{dyn}=\varphi$.

But we will show in the next subsections that the situation may be slightly more complex.

### 3.4 Effect of a half cycle

Some calculations with the Drücker-Prager law have been performed starting rotating in one way during few degrees, *i.e.* less than 5°, then rotating in the reverse direction till breaking is provoked. It turns out that the maximum angle of repose $\theta_{M,h}$ which is obtained using this procedure is always larger than $\varphi$, but is also much less than the maximum angle of repose measured without this first half cycle. This lets predict a way to measure the fiction angle more accurately. This result shows a strong history dependence of $\theta_M$ .

It may be understood in the following way: as shown in Fig. 6, during the first increase of $\theta$ during the first half cycle, the medium evolves plastically in such a way that it becomes harder for this kind of rotation; then, the stress remains in the elastic domain during the $\theta$ decrease till it reaches the starting angle $\theta_o$=25°; this irreversibility makes the pile less adaptive.

So this result may be quite important if it is confirmed, since it may explain why most yielding occurs near $\theta_M=\varphi$. However, it has to be confirmed by more numerical simulations.

### 3.5 Effect of dilatancy

Different flow rules have been used in these computations; they were associated with the two yield criteria, *i.e.* Drücker-Prager *vs.* Mohr-Coulomb. So some of these simulations were using associate flow rules, others not, so that the effect of dilatancy has been tested too. It turns out that the breaking threshold $\theta_M$ which have been obtained depend much more on the building process and on the yield criterion. The role of the dilatancy is particularly important in the case of Drücker-Prager law. This has been already explained in paragraph 3.1.

Anyway, it is worth repeating that $\theta_M=\varphi$ has been always obtained when the Mohr-Coulomb yield criterion has been used, and $\theta_M>\varphi$ with the Drücker-Prager one.

### *3.6 Sensitivity of the calculation to details*

Owing to the arguments developed in the preceding subsections, one shall then conclude that the calculated predicted behaviour is sensitive to the details of the pile construction, of the pile history and to the stress yield criterion. However, they seem to be less sensitive to the flow rule.

## 4. Conclusion

These simulations use a basic soil mechanics model, without cohesion. They start with a symmetrical triangular pile with a free surface inclined at an angle $\theta_o$ smaller than $\varphi$. The pile is then rotated resulting in the increase of the slope $\theta$ on one side and the decrease of the other slope hence.

The calculations demonstrate that the results depend on the rheological law which is chosen . When one choses a Mohr-Coulomb criterion, the slope of the pile can not overpass the friction angle.

However this is different with a Drücker-Prager law, since the calculations have demonstrated in this case that:

(i) stable solutions do exist for a free surface more inclined than the friction angle $\varphi$; these solutions are at least marginally stable,

(ii) the equations do not converge for angles $\theta>\varphi$; and $\theta-\varphi$ can be as large as $\delta\theta_{aval}=10°$,

(iii) cyclic loading whose amplitude $\Delta\theta<\varphi-\theta_o$ limits the slope range achievable so that $\delta\theta_{aval}$ can be reduced by the change of stress state,

(iv) if a first half cycle is performed with amplitude $\Delta\theta>\varphi-\theta_o$, then the system breaks often at $\theta=\varphi$ for the other slope in the case of an instantaneous pile construction.

(v) results depend on material stress state and history.

(vi) the flow rule is also important and can modify the value found for the maximum slope. This is normal, since evolution of stress and strain are coupled together via the flow rule in a plastic modelling.

(vii) These results can be and have been understood from a theoretical approach, since the equivalent of the friction angle is sensitive to the value of the intermediate stress in a Drücker-Prager modelling.

Owing to this one can ask whether the Drücker-Prager modelling has some physical meaning or not. However, as far as the calculation of the stress field in an inclined pile with a slope smaller than the friction angle finds in general that the intermediate stress lays rather near to the smaller one, it is difficult to suppose that the two stress directions of $\sigma_2$ and $\sigma_3$ play complete different roles. But this should be supposed when one wants to apply the Mohr-Coulomb criterion instead of the Drücker-Prager one, since Mohr-Coulomb criterion assumes the symmetry breaking between $\sigma_2$ and $\sigma_3$. Conversely, it means then that the Drücker-Prager criterion shall be used preferentially in this range of stress, which let predicts some uncertainty of the friction angle definition in turn.

This work demonstrates then:

(i) that rotation of the slope is not always a strong perturbation which tests truly the stability of the slope,

(ii) that piles obeying classical constitutive models, such as the Drücker-Prager law, *i.e.* solid friction $\varphi$ without cohesion, can exhibit a maximum free-surface slope which is not well defined and can be larger than the definition of $\varphi$,

(iii) in these cases the solutions which are found larger than $\varphi$ are marginally stable so that avalanche process is included in classical soil mechanics. The stability of these solutions will depend on the perturbations which are applied on the pile.

In conclusion there is probably no need to introduce a dynamic friction angle different from the static one to understand the mechanics of avalanche; the static problem is already rather complicated.

However, this work settles a lot of questions: it proves that solutions proposed by computer simulations for engineering works can be often only marginally stable; this means that they can potentially fail. So, it will be important in the future to develop a probabilistic approach which can take into account this failure possibility, and which determines the type of perturbations which are the most sensitive.

The other question could be the necessity of the introduction of another criterion to detect instability.

Indeed other marginally stable situations are known already in soil mechanics: for instance marginally stable situations are found during simple triaxial undrained compression of loose samples or during compression of dense drained samples. So they are really achievable experimentally, and their failure probability shall be investigated then incorporated in the mechanical description.

At last, this work has emphasised that the Drücker-Prager law leads to a behaviour different from the Mohr-Coulomb one. It has also argued that the Mohr-Coulomb law assumes a breaking of symmetry of the role plaid by the two minor stress directions $\sigma_2$ and $\sigma_3$; this is rather not physical when $\sigma_2 \approx \sigma_3$ , so that it is better to use the Drücker-Prager law in this case. However it is probably much better to use the Mohr-Coulomb criterion as soon as $\sigma_2 > \sigma_3$, since experiments show unambiguously that the friction angle is a well defined quantity. So, it is also important to study how the behaviour starts from a Drücker-Prager law at when $\sigma_2 = \sigma_3$ and changes gradually to a Mohr-Coulomb one at when $\sigma_2 > \sigma_3$ . This may help understanding results on strain localisation in triaxial tests with axial symmetry on granular material.

So, the only "new" facts are (i) that such metastable solutions can also occur for complex objects, such as a pile and (ii) that slope rotation is not always a critical perturbation for the stability of the slope.

*Acknowledgements:* CNES is thanked for partial funding.

## References

[1] J.P. Bouchaud, M. Cates, J. Ravi Prakash, S.F. Edwards, *J. Phys. France* **4**, 1383 (1994)

[2] P. Evesque, D. Fargeix, P. Habib, M.P. Luong & P. Porion, "Pile Density is a control parameter of sand avalanches", *Phys. Rev.* **E 47**, 2326 (1993)

[3] P. Evesque, "Can we define a unique friction coefficient forr non cohesive granular material? A temptative answer from the point of view of sand avalanche experiment", in *Proceedings of the "Pierre Beghin" international workshop on rapid gravitational mass movements*, (L. Buisson & G. Brugnot eds), (Antony, France: Cemagref-Dicova, 1993)

[4] P. Porion, *Frottement solide et avalanches dans les matériaux granulaires*. PhD. Thesis, Université de Lille 3, France, (1994)

[5] P. Porion & P. Evesque, "Density is a controlling parameter of sandpile avalanches", in *Powders & Grainss 93* (C. Thornton ed.), *(*Rotterdam: Balkema, 1993), pp. 327-332

[6] A. Modaressi, S. Boufellouh, S. & P. Evesque, 1"Modelling of stress distribution in granular pile : comparison with centrifuge experiments". *Chaos* **9**, 523-543, (1999)

[7] A. Schinner, H.-G. Mattutis, T. Akiyama, J. Aoki, S. Takahashi, K.M. Aoki & K. Kassner, "History-dependence structure in granular piles", in Powders & Grains 2001, Y. Kishino ed., (Balkema, Rotterdam, 2001), pp. 499-502

[8] M. Pastor, M. Quecedo, "A Patch Test for Mesh Alignment Effects in Localized Failure". *Communications in numerical methods in engineering* DEC 01 1995 v 11 n 12.

[9] F. Cantelaube & J.D. Goddard, "Elastoplastic arching in 2D granular heap"., in *Powders & Grainss 97*, (R.P. Behringer & J.T. Jenkins eds.), 231-234, (Rotterdam, Balkema, 1997.)

[10] B. Halphen & J. Salençon, Elasto-plasticité, Presses des Ponts & Chaussées, (1985)

[11] P.V. Lade & J.M. Duncan, "Cubical triaxial tests on cohesionless soil", *J. Geot. Engnq. Div. (*ASCE*)* **99**, n°5M10, pp. 793-812, (1973)

The electronic arXiv.org version of this paper has been settled during a stay at the Kavli Institute of Theoretical Physics of the University of California at Santa Barbara (KITP-UCSB), in june 2005, supported in part by the National Science Fundation under Grant n° PHY99-07949.

*Poudres & Grains* can be found at :
http://www.mssmat.ecp.fr/rubrique.php3?id_rubrique=402

The electronic arXiv.org version of this paper has been settled during a stay at the Kavli Institute of Theoretical Physics of the University of California at Santa Barbara (KITP-UCSB), in june 2005, supported in part by the National Science Fundation under Grant n° PHY99-07949.

*Poudres & Grains* can be found at :
http://www.mssmat.ecp.fr/rubrique.php3?id_rubrique=402